\documentclass[12pt]{article}
\usepackage{epsfig}
\def\be{\begin{equation}}
\def\ee{\end{equation}}
\def\bea{\begin{eqnarray}}
\def\eea{\end{eqnarray}}
\usepackage{graphicx}

\catcode`\@=11
\def\lsim{\mathrel{\mathpalette\@versim<}}
\def\gsim{\mathrel{\mathpalette\@versim>}}
\def\@versim#1#2{\vcenter{\offinterlineskip
\ialign{$\m@th#1\hfil##\hfil$\crcr#2\crcr\sim\crcr } }}
\catcode`\@=12

\catcode`@=12
\begin{document}
\thispagestyle{empty}
\begin{flushright}
UCRHEP-T554\\
July 2015\
\end{flushright}
\vspace{0.6in}
\begin{center}
{\LARGE \bf Gauge $B-L$ Model with\\ Residual $Z_3$ Symmetry\\}
\vspace{1.0in}
{\bf Ernest Ma$^1$, Nicholas Pollard$^1$, Rahul Srivastava$^2$, and 
Mohammadreza Zakeri$^1$\\}
\vspace{0.5in}
{\sl $^1$Department of Physics and Astronomy,\\}
{\sl University of California, 
Riverside, California 92521, USA\\}
\vspace{0.2in}
{\sl $^2$The Institute of Mathematical Sciences, Chennai 600 113, India}
\end{center}
\vspace{1.0in}

\begin{abstract}\
We study a gauge $B-L$ extension of the standard model of quarks and leptons 
with unconventional charges for the singlet right-handed neutrinos, and 
extra singlet scalars, such that a residual $Z_3$ symmetry remains after 
the spontaneous breaking of $B-L$.  We discuss the phenomenological 
consequences of this scenario, including the possibility of long-lived 
self-interacting dark matter and $Z'$ collider signatures. 
\end{abstract}

\newpage
\baselineskip 24pt

Lepton number $L$ is a familiar concept.  It is usually defined as a global 
U(1) symmetry, under which the leptons of the standard model (SM), i.e. 
$e,\mu,\tau$ together with their neutrinos $\nu_e,\nu_\mu,\nu_\tau$ have 
$L = 1$, and all other SM particles have $L = 0$.  In the case of nonzero 
Majorana neutrino masses, this continuous symmetry is broken to a discrete 
$Z_2$ symmetry, i.e. $(-1)^L$ or lepton parity.  In this paper, we consider 
a gauge $B-L$ extension of the SM, such that a residual $Z_3$ symmetry 
remains after the spontaneous breaking of $B-L$.  This is then a realization 
of the unusual notion of $Z_3$ lepton symmetry.  It has specific 
phenomenological consequences, including the possibility of a long-lived 
particle as a dark-matter candidate.

The conventional treatment of gauge $B-L$ has three right-handed singlet 
neutrinos $\nu_{R1},\nu_{R2},\nu_{R3}$ transforming as $-1,-1,-1$ under $B-L$. 
It is well-known that this assignment satisfies  all the anomaly-free 
conditions for $U(1)_{B-L}$.  However, another assignment~\cite{mp09}
\begin{equation}
\nu_{R1}, \nu_{R2}, \nu_{R3} \sim 5,-4,-4
\end{equation}
works as well, because
\begin{equation}
5-4-4 = -3, ~~~ (5)^3 - (4)^3 - (4)^3 = -3.
\end{equation}
To obtain a realistic model with this assignment, it was recently 
proposed~\cite{ms15} that three additional neutral singlet Dirac fermions 
$N_{1,2,3}$ be added with $B-L = -1$, together with a singlet scalar 
$\chi_3$ with $B-L = 3$.  Consequently, the tree-level Yukawa couplings 
$\bar{\nu}_L N_R \bar{\phi}^0$ and $\bar{N}_L \nu_{R2} \chi_3$, 
$\bar{N}_L \nu_{R3} \chi_3$ are allowed, 
where $\Phi = (\phi^+,\phi^0)$ is the one Higgs doublet of the SM.  Together 
with the invariant $\bar{N}_L N_R$ mass terms, the $6 \times 5$ neutrino 
mass matrix linking $(\bar{\nu}_L,\bar{N}_L)$ to $(\nu_R,N_R)$ is of the 
form
\begin{equation}
{\cal M}_{\nu N} = \pmatrix{ 0 & {\cal M}_0 \cr {\cal M}_3 & {\cal M}_N},
\end{equation}
where ${\cal M}_0$ and ${\cal M}_N$ are $3 \times 3$ mass matrices and 
${\cal M}_3$ is $3 \times 2$ because $\nu_{R1}$ has no tree-level Yukawa 
coupling.  This means that one linear combination of $\nu_L$ is massless.
Of course, if the dimension-five term $\bar{\nu}_{R1} N_L \chi_3^2$ also 
exists, then ${\cal M}_3$ is $3 \times 3$ and ${\cal M}_{\nu N}$ is 
$6 \times 6$.  

The form of ${\cal M}_{\nu N}$ allows nonzero seesaw Dirac neutrino masses 
for $\nu$~\cite{rs84}, i.e.
\begin{equation}
{\cal M}_\nu \simeq {\cal M}_0 {\cal M}_N^{-1} {\cal M}_3.
\end{equation}
Without the implementation of a flavor symmetry, any $3 \times 3 ~{\cal M}_\nu$ 
is possible.  Although the gauge $B-L$ is broken, a residual global $L$ 
symmetry remains in this model with $\nu, l, N$ all having $L=1$.  Because 
the pairing of any two neutral fermions of the same chirality always results 
in a nonzero $B-L$ charge not divisible by 3 units in this model, it is 
impossible to construct an operator of any dimension for a Majorana mass 
term which violates $B-L$.  Hence the neutrinos are indeed exactly Dirac.

We now add two more scalar singlets: $\chi_2$ with $B-L = 2$ and $\chi_6$ 
with $B-L = -6$.  The important new terms in the Lagrangian are
\begin{equation}
\bar{N}_L \nu_{R1} \chi_6, ~~~ \chi_2 N_L N_L, ~~~ \chi_2 N_R N_R, ~~~ 
\chi_2^3 \chi_6, ~~~ \chi_3^2 \chi_6.
\end{equation}
Now $B-L$ is broken by $\langle \chi_3 \rangle = u_3$ as well as 
$\langle \chi_6 \rangle = u_6$, and all neutrinos become massive. 
The cubic term $\chi_2^3$ implies that a $Z_3$ residual symmetry remains, 
such that $\chi_2$ and all leptons transform as 
$\omega = \exp(2 \pi i/3)$ under $Z_3$.  This is thus the first example of 
a lepton symmetry which is not $Z_2$ (for Majorana neutrinos), nor $U(1)$ 
or $Z_4$~\cite{hr13,h13} (for Dirac neutrinos).  Note that $Z_3$ 
is also sufficient to guarantee that all the neutrinos remain Dirac.  

Although there is no stabilizing symmetry here for dark matter, $\chi_2$ 
has very small couplings to two neutrinos through the Yukawa 
terms of Eq.~(5) from the mixing implied by Eq.~(3).  This means that 
$\chi_2$ may have a long enough lifetime to be suitable for dark matter, 
as shown below.

Consider for simplicity the coupling of $\chi_2$ to just one $N$, with the 
interaction
\begin{equation}
{\cal L}_{int} = {1 \over 2} f_L \chi_2 N_L N_L + {1 \over 2} f_R \chi_2 
N_R N_R + H.c.
\end{equation}
Let the $\nu_L - N_L$ mixing be $\zeta_0 = m_0/m_N$ and $\nu_R - N_R$ 
mixing be $\zeta_3 = m_3/m_N$, then the decay rate of $\chi_2$ is 
\begin{equation}
\Gamma (\chi_2 \to \bar{\nu} \bar{\nu}) = {m_\chi \over 32 \pi} (f_L^2 
\zeta_0^4 + f_R^2 \zeta_3^4).
\end{equation}
If we set this equal to the age of the Universe ($13.75 \times 10^9$ years), 
and assuming $m_\chi = 100$ GeV, $f_L = f_R$ and $\zeta_0 = \zeta_3$, then 
$f \zeta^2 = 8.75 \times 10^{-22}$.  Hence
\begin{equation}
\sqrt{f} \zeta << 3 \times 10^{-11}
\end{equation}
would guarantee the stability of $\chi_2$ to the present day, and allow 
it to be a dark-matter candidate.  This sets the scale of $m_N$ at about 
$10^{13}$ GeV, which is also the usual mass scale for the heavy Majorana 
singlet neutrino in the canonical seesaw mechanism.

In this model, there is of course a gauge boson $Z'$ which couples to $B-L$. 
Its production at the Large Hadron Collider (LHC) is due to its couplings  
to quarks.  Once produced, it decays into quarks and leptons. 
In the conventional $B-L$ assignment for $\nu_R$, its branching fractions 
to quarks, charged leptons, and neutrinos are 1/4, 3/8, and 3/8 respectively. 
In this model, the $\nu_R$ charges are $(5,-4,-4)$, hence their resulting 
partial widths are very large.  Assuming that $Z'$ decays also into $\chi_2$, 
the respective branching fractions into quarks, charged leptons, neutrinos, 
and $\chi_2$ as dark matter are then 1/18, 1/12, 5/6, and 1/36.  This means 
$Z'$ has an 86\% invisible width.  Using the production of $Z'$ via 
$u \bar{u}$ and $d \bar{d}$ initial states at the LHC and its decay into 
$e^- e^+$ or $\mu^- \mu^+$ as signature, the current bound on $m_{Z'}$ 
assuming $g' = g$, i.e. the $SU(2)_L$ gauge coupling of the SM, is 
about 3 TeV, based on recent LHC data~\cite{atlas14,cms15}.  However, 
because the branching fraction into $l^- l^+$ is reduced by a factor of 
2/9 in our $B-L$ model, this bound is reduced to about 2.5 TeV, again 
for $g' = g$.

Since $\chi_2$ interacts with nuclei through $Z'$, there is also a 
significant constraint from dark-matter direct-search experiments. 
The cross section per nucleon is given by
\begin{equation}
\sigma_0 = {1 \over \pi} \left( {m_\chi m_n \over m_\chi + A m_n} \right)^2 
\left( {2 {g'}^2 \over m^2_{Z'}} \right)^2,
\end{equation}
where $A$ is the number of nucleons in the target and $m_n$ is the nucleon 
mass.  Consider for example $m_\chi = 100$ GeV, then  $\sigma_0 < 1.25 \times  
10^{-45}~{\rm cm}^2$ from the recent LUX data~\cite{lux14}.  This 
implies $m_{Z'}/g' > 16.2$ TeV, as shown in Fig.~1.  
\begin{figure}[htb]
\vspace*{0.2cm}
\hspace*{1.0cm}
\includegraphics[scale=1.0]{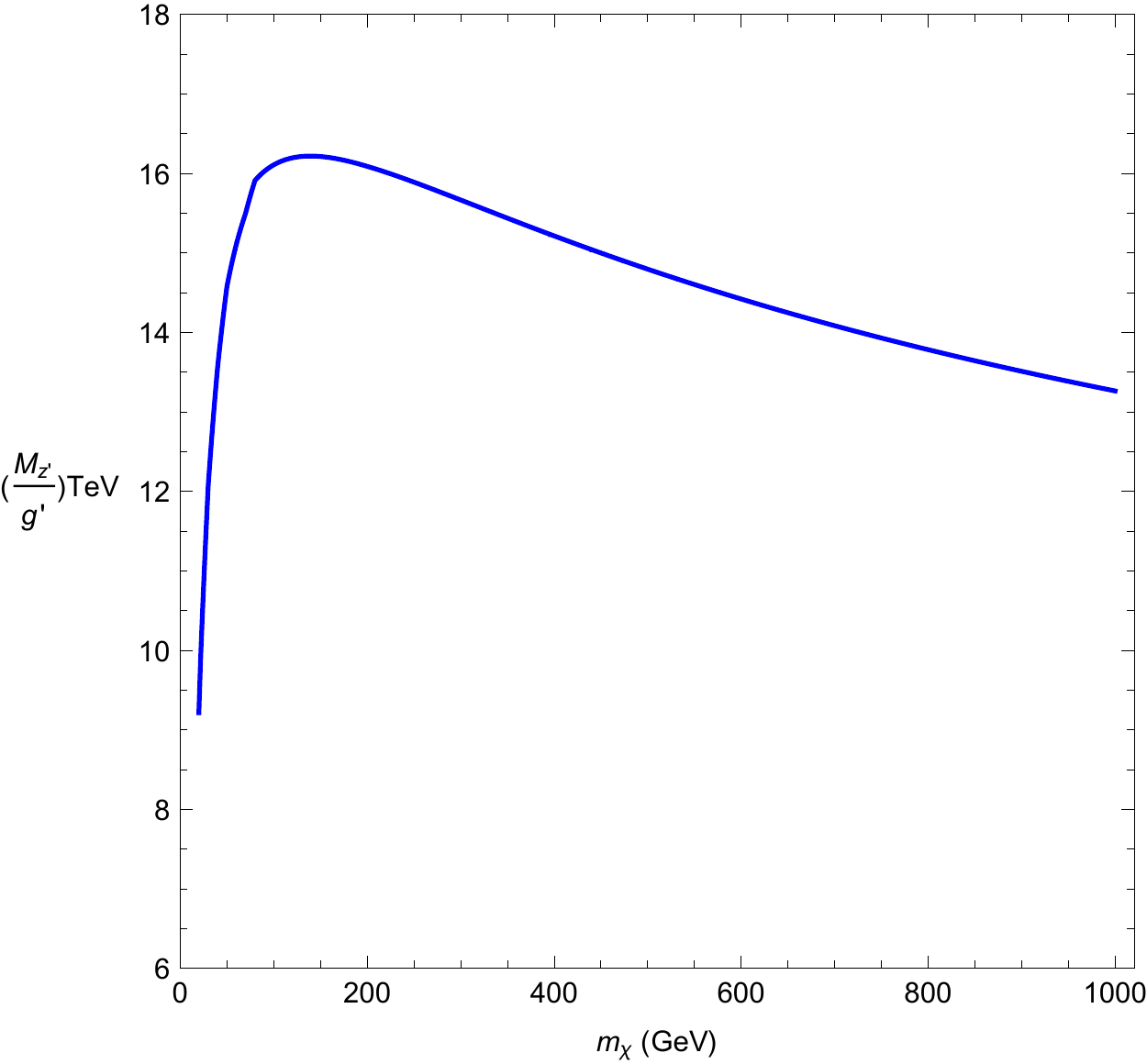}
\caption{Lower bound on $m_{Z'}/g'$ versus $m_\chi$ from LUX data.}
\end{figure}
If $g' = g$, then 
$m_{Z'} > 10.6$ TeV.  
This limit is thus much more severe than the LHC bound of 2.5 TeV. 
If $g' < g$, then both the LHC and LUX bounds on $m_{Z'}$ are relaxed.  
However, it also means that it is unlikely that $Z'$ would be discovered 
at the LHC even with the 14 TeV run.

Consider now the annihilation cross section of $\chi_2 \chi_2^*$ for 
obtaining its thermal relic abundance.  The process $\chi_2 \chi_2^* \to 
Z' \to$ SM particles is $p$-wave suppressed and is unlikely to be strong 
enough for this purpose.  We may then consider the well-studied process 
$\chi_2 \chi_2^* \to h \to$ SM particles, where $h$ is the SM Higgs 
boson.  If this is assumed to account for all of the dark-matter relic 
abundance of the Universe, then it has recently been shown~\cite{fpu15} 
that the required strength of this interaction is in conflict with LUX data 
except for a small region near $m_\chi = m_h/2$.

\begin{figure}[htb]
\vspace*{-3cm}
\hspace*{-3cm}
\includegraphics[scale=1.0]{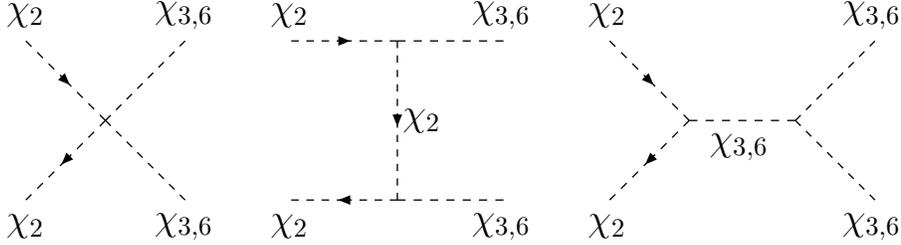}
\vspace*{-21.5cm}
\caption{$\chi_2 \chi_2^\dagger$ annihilation to $\chi_{3,6}$ final states.}
\end{figure}
In this paper, we will consider the following alternative scenario. We 
assume that the $h \chi_2 \chi_2^*$ interaction is negligible, so that neither 
Higgs nor $Z'$ exchange is important for $\chi_2 \chi_2^*$ annihilation. 
Instead we invoke the new interactions of Fig.~2. 
Since $\chi_{3,6}$ may interact freely with $h$, thermal equilibrium is 
maintained with the other SM particles.  This scenario requires of course 
that $m_\chi$ to be greater than at least one physical 
mass eigenvalue in the $\chi_{3,6}$ sector.

To summarize, $\chi_2 \sim \omega$ under $Z_3$ and decays into two 
antineutrinos, but its lifetime is much longer than the age of the 
Universe.  It is thus an example of $Z_3$ dark 
matter~\cite{m08,bkpr13,kt14,gkko15,m15}.  It is also different from 
previous $Z_2$ proposals~\cite{sms14,ss15} based on Ref.~\cite{mp09}.  
It has significant elastic 
interactions with nuclei through $Z'$ and Higgs exchange and may be discovered 
in direct-search experiments.  On the other hand, its relic abundance 
is determined not by  $Z'$ or Higgs interactions, but by its annihilation 
to other scalars 
of this model which maintain thermal equilibrium with the SM particles 
through the SM Higgs boson.  Note that this is also the mechanism used 
in a recently proposed model of vector dark matter~\cite{fmz15}.

We now discuss the details of the scalar sector of this model.  Consider 
the scalar potential
\begin{eqnarray}
V &=& -\mu_0^2 (\Phi^\dagger \Phi) + m_2^2 (\chi_2^* \chi_2) - \mu_3^2 
(\chi_3^* \chi_3) - \mu_6^2 (\chi_6^* \chi_6) \nonumber \\ 
&+& {1 \over 2} \lambda_0 (\Phi^\dagger \Phi)^2 + {1 \over 2} \lambda_2 
(\chi_2^* \chi_2)^2 + {1 \over 2} \lambda_3 (\chi_3^* \chi_3)^2 + 
{1 \over 2} \lambda_6 (\chi_6^* \chi_6)^2 + \lambda_{02} (\chi_2^* \chi_2) 
(\Phi^\dagger \Phi) \nonumber \\ 
&+& \lambda_{03} (\chi_3^* \chi_3) (\Phi^\dagger \Phi) 
+ \lambda_{06} (\chi_6^* \chi_6) (\Phi^\dagger \Phi) + \lambda_{23} 
(\chi_2^* \chi_2) (\chi_3^* \chi_3) + \lambda_{26} (\chi_2^* \chi_2) 
(\chi_6^* \chi_6) \nonumber \\
&+& \lambda_{36} (\chi_3^* \chi_3) (\chi_6^* \chi_6) 
+ [ {1 \over 2} f_{36} (\chi_3^2 \chi_6) + {\rm H.c.}] 
+ [ {1 \over 6} \lambda'_{26} (\chi_2^3 \chi_6) + {\rm H.c.}].
\end{eqnarray}
Let $\langle \phi^0 \rangle = v$,  $\langle \chi_3 \rangle = u_3$,  
$\langle \chi_6 \rangle = u_6$,  then the minimum of $V$ is determined by
\begin{eqnarray}
\mu_0^2 &=& \lambda_0 v^2 + \lambda_{03} u_3^2 + \lambda_{06} u_6^2, \\ 
\mu_3^2 &=& \lambda_3 u_3^2 + \lambda_{03} v^2 + \lambda_{36} u_6^2 
+ f_{36} u_6, \\ 
\mu_6^2 &=& \lambda_6 u_6^2 + \lambda_{06} v^2 + \lambda_{36} u_3^2 
+ {f_{36} u_3^2 \over 2 u_6}. 
\end{eqnarray}
There is one dark-matter scalar boson $\chi_2$ with mass given by
\begin{equation}
m_\chi^2 = m_2^2 + \lambda_{02} v^2 + \lambda_{23} u_3^2 + \lambda_{26} 
u_6^2.
\end{equation}
There is one physical pseudoscalar boson
\begin{equation}
A = \sqrt{2} Im(2 u_6 \chi_3 + u_3 \chi_6)/\sqrt{u_3^2 + 4 u_6^2}
\end{equation}
with mass given by
\begin{equation}
m_A^2 = - f_{36} (u_3^2 + 4 u_6^2)/2u_6.
\end{equation}
There are three physical scalar bosons spanning the basis $[h, \sqrt{2} 
Re(\chi_3), \sqrt{2} Re(\chi_6)]$, with $3 \times 3$ mass-squared matrix 
given by
\begin{equation}
M^2 = \pmatrix{ 2 \lambda_0 v^2 & 2 \lambda_{03} u_3 v & 2 \lambda_{06} u_6 v 
\cr  2 \lambda_{03} u_3 v & 2 \lambda_3 u_3^2 & 2 \lambda_{36} u_3 u_6 + 
f_{36} u_3 \cr 2 \lambda_{06} u_6 v & 2 \lambda_{36} u_3 u_6 + 
f_{36} u_3 & 2 \lambda_6 u_6^2 - f_{36} u_3^2/2 u_6}.
\end{equation}
For illustration, we consider the special case $\lambda_{03} = \lambda_{06} = 0$, 
so that $h$ decouples from $\chi_{3,6}$.  It then becomes identical to that 
of the SM, and may be identified with the 125 GeV particle 
discovered~\cite{atlas12,cms12} at the LHC.  We now look for a solution with
\begin{eqnarray}
S &=& \sqrt{2} Re(-u_3 \chi_3 + 2u_6 \chi_6)/\sqrt{u_3^2 + 4 u_6^2}, \\ 
S' &=& \sqrt{2} Re(2 u_6 \chi_3 + u_3 \chi_6)/\sqrt{u_3^2 + 4 u_6^2},
\end{eqnarray}
as mass eigenstates.  This is easily accomplished for example with
\begin{equation}
u_3 = 2 u_6, ~~~~~ 4 \lambda_3 = \lambda_6 - f_{36}/u_6.
\end{equation}
In this case,
\begin{eqnarray}
S = -Re \chi_3 + Re \chi_6, && m^2_{S} = 2 \lambda_6 u_6^2 - 
4 \lambda_{36} u_6^2 - 4 f_{36} u_6, \\ 
S' = Re \chi_3 + Re \chi_6, && m^2_{S'} = 2 \lambda_6 u_6^2 + 
4 \lambda_{36} u_6^2, \\ 
A = Im \chi_3 + Im \chi_6, && m^2_A = -4 f_{36} u_6, \\ 
&& m_{Z'} = 12 g' u_6.
\end{eqnarray}
The couplings of $\chi_2 \chi_2^*$ to $S$ and $S'$ are given by
\begin{equation}
\chi_2 \chi_2^* [u_6 (\lambda_{26} - 2 \lambda_{23}) S + 
u_6 (\lambda_{26} + 2 \lambda_{23}) S'].
\end{equation} 
Since $S$ plays the same role in breaking $B-L$ as the Higgs boson $h$ does 
in breaking $SU(2)_L \times U(1)_Y$, it is expected to be massive of order 
$\sqrt{u_3^2 + 4 u_6^2} = 2\sqrt{2} u_6$.  This allows $m_{S'}$ to be adjusted 
to be very small, then it may serve as a light scalar mediator for $\chi_2$ 
as self-interacting dark matter~\cite{t14}.  For $m_{S'} \simeq 0$, we need 
$\lambda_{36} = -\lambda_6/2$.  In that case, using Eq.~(20), we find
\begin{equation}
m_S^2 = 16 \lambda_3 u_6^2, ~~~ m_A^2 = m_S^2 - 4 \lambda_6 u_6^2.
\end{equation}

We assume that the relic density of $\chi_2$ is dominated by the 
$\chi_2 \chi_2^*$ annihilation to $S' S'$.  For illustration, we set to 
zero the $\chi_2 \chi_2^* S' S'$ coupling, i.e. $\lambda_{23} + \lambda_{26} 
= 0$, as well as the $S S' S'$ coupling, i.e. $-12 \lambda_3 + 
6 \lambda_6 + 2 \lambda_{36} - f_{36}/u_6 = 0$.  This implies $\lambda_3 = 
\lambda_6/2$ so that the $S' S' S'$ coupling is also zero and $m_A^2 = 
m_S^2/2$.   This choice of parameters means that only the middle diagram 
of Fig.~2 contributes to the $\chi_2 \chi_2^*$ annihilation cross section 
with
\begin{equation}
\sigma \times v_{rel} = {1 \over 64 \pi m_\chi^2} \left| {\lambda_{26}^2 
u_6^2 \over m_\chi^2} \right|^2.
\end{equation}
Equating this to the optimal value~\cite{sdb12} of $4.4 \times 
10^{-26}~{\rm cm}^3~{\rm s}^{-1}$ for the correct dark-matter relic density 
of the Universe, we find for $m_\chi = 100$ GeV 
\begin{equation}
\lambda_{26} = 0.0295 \left( 
{1~{\rm TeV} \over u_6} \right).
\end{equation}
We assume of course that $m_A > 2 m_\chi$.

For $S'$ to be in thermal equilibrium with the SM particles, we need to 
have nonzero values of $\lambda_{03}$ and $\lambda_{06}$.  This is possible 
in our chosen parameter space if $2\lambda_{03} + \lambda_{06}  \simeq0$, 
so that the $S' h$ mixing is very small and yet the $S' S' h$ coupling 
$\lambda_{06} v/4 \sqrt{2}$ and $S' S' h h$ coupling $\lambda_{06}/16$ 
may be significant.  Note that the $S h$ mixing is now fixed at 
$(\lambda_{06}/\lambda_6)(v/2 \sqrt{2} u_6)$ which may yet be suitably 
suppressed for $h$ to be essentially the one Higgs boson of the SM.
The $h \to S' S'$ decay width is given by
\begin{equation}
\Gamma (h \to S' S') = {\lambda_{06}^2 v^2 \over 256 \pi m_h} = 
\left( {\lambda_{06} \over 0.04} \right)^2 0.5~{\rm MeV}.
\end{equation}
It is invisible at the LHC because $S'$ decays slowly 
to $e^- e^+$ only through its mixing with $h$, if $m_{S'} \sim  10$ MeV 
for $S'$ as a light mediator for the self-interacting dark matter $\chi_2$.

In conclusion, we have considered the unusual case of a gauge $B-L$ symmetry 
which is spontaneously broken to $Z_3$ lepton number.  Neutrinos are 
Dirac fermions transforming as $\omega = \exp (2 \pi i/3)$ under $Z_3$. 
A complex neutral scalar $\chi_2$ exists which also transforms as $\omega$. 
It is not absolutely stable, but decays to two antineutrinos with a 
lifetime much greater than that of the Universe.  It is thus an example 
of $Z_3$ dark matter.  In addition to the one Higgs boson $h$ of the SM, 
there are three neutral scalars $S, S', A$ and one heavy vector gauge 
boson $Z'$.  From direct-search experiments, $m_{Z'}/g'$ is constrained 
to be very large, thus making it impossible to discover $Z'$ at the 
LHC even with the current run.  The relic abundance of $\chi_2$ is 
determined by its annihilation into $S'$ which is a candidate for 
the light mediator by which $\chi_2$ obtains its long-range 
self-interaction.

\medskip

This work is supported in part 
by the U.~S.~Department of Energy under Grant No.~DE-SC0008541.

\bibliographystyle{unsrt}

\end{document}